# Violation of the Wiedemann-Franz law as an evidence of the pseudogap in the iron-based superconductor Ba(Fe$_{1-x}$Co$_x$)$_2$As$_2$

Marcin Matusiak[1,*] and Thomas Wolf[2]

1. *Institute of Low Temperature and Structure Research, Polish Academy of Sciences, ul. Okolna 2, 50-422 Wroclaw, Poland*

2. *Institute of Solid State Physics (IFP), Karlsruhe Institute of Technology, D-76021, Karlsruhe, Germany*

**PACS**

7 4.70.Xa, 74.25.fc, 72.15.Jf

Longitudinal and transverse transport coefficients of the Ba(Fe$_{1-x}$Co$_x$)$_2$As$_2$ single crystals with $x$ = 0, 0.045, 0.06 and 0.244 were measured in the temperature range 1.4 – 300 K and in magnetic fields up to 12.5 T. The resulting data were used to determine the temperature dependence of the Hall Lorenz number ($L_{xy}$) and its evolution with doping. $L_{xy}$ is defined by the electronic contributions to the thermal and electrical conductivities and it is found to differ from its canonical behavior. This shows the emergence of a pseudogap in samples at intermediate doping.

**Introduction**

Since the discovery of a pseudogap (PG) in high-$T_c$ superconductors, it has become one of the most extensively studied phenomena in condensed matter physics. Nevertheless, the subject is far from being elucidated. Even the basic question, whether the pseudogap is a precursor of the superconducting state [1–3] or these two are independent/competitive phenomena [4–7] remains controversial. In the meantime, there have been published reports suggesting that the pseudogap may be a common feature of various families of unconventional superconductors, including heavy fermion [8] and topological [9] ones. There is growing experimental evidence that the pseudogap is also present in iron-based superconductors. The pseudogap was reported to occur in the "11" iron chalcogenide $Fe_{1+d}Te_{1-x}Se_x$ [10], "1111" Fe-pnictides $LaFeAs(O_{1-x}F_x)$ [11,12] and $SmFeAs(O_{1-x}F_x)$ [13, 14], as well as "122" hole-doped $Ba_{1-x}K_xFe_2As_2$ [15,16] and electron-doped $Ba(Fe_{1-x}Co_x)_2As_2$ [17,18]. On the other hand, some studies provide the opposite evidence. For example, scanning tunneling spectroscopy measurements performed on $Ba(Fe_{1-x}Co_x)_2As_2$ showed no sign of PG [19].

In the present letter we investigate properties of a series of $Ba(Fe_{1-x}Co_x)_2As_2$ single crystals by means of the Righi-Leduc effect (also called the thermal Hall effect). The Righi-Leduc coefficient combined with data on the resistivity, thermal conductivity and Hall coefficient gives a unique opportunity to investigate purely electronic heat transport and the relation between the thermal and electrical conductivities, known as the Wiedemann-Franz law (WFL). A depletion of the electronic density of states at the Fermi level is expected to cause violation of this law [20–25] and we investigate such a scenario in the iron pnictide $Ba(Fe_{1-x}Co_x)_2As_2$.

**Experimental**

The $Ba(Fe_{1-x}Co_x)_2As_2$ crystals we used were grown from self-flux in glassy carbon or alumina crucibles. Particularly low cooling rates of 0.20 to 0.61°C/h were applied to minimize the amount of flux inclusions and crystal defects [26]. The composition of the crystals were determined by energy dispersive x-ray spectroscopy.

All transport coefficients were measured along the *ab*-plane of a $Ba(Fe_{1-x}Co_x)_2As_2$ crystal with a magnetic field (*B*) applied parallel to the *c* axis. The electrical resistivity ($\rho$) was determined using a four-point technique. The Hall coefficient ($R_H$) measurements were performed by a standard method, where current and magnetic field directions were reversed several times to exclude any influence of the asymmetric position of the Hall contacts and detrimental electromotive forces. The longitudinal ($\kappa_{xx}$) and transverse ($\kappa_{xy}$) thermal conductivities were measured in a single experiment, where respective temperature gradients ($\nabla_x T$ and $\nabla_y T$) were determined with Chromel-Constantan (Type E) thermocouples that have been calibrated in a set of magnetic fields. Typically $\nabla_x T$ and $\nabla_y T$ were of order 1 K/mm and 10 mK/mm, respectively. Measurements were repeated in various magnetic fields between -12.5 T and +12.5 T in order to separate field-symmetric and -antisymmetric components of the signal.

**Results and Discussion**

To provide a comprehensive view of the electronic system in Ba(Fe$_{1-x}$Co$_x$)$_2$As$_2$ we prepared a series of single crystals having superconducting (SC) and spin density wave (SDW) phases. Namely, we studied the parent compound BaFe$_2$As$_2$ (denoted as Co0) along with three cobalt doped samples with $x$ = 0.045 (Co4), 0.06 (Co6) and 0.244 (Co24). Table 1 collects transition temperatures to the SDW ($T_{SDW}$) and to the SC ($T_c$) state determined from resistivity data presented in Fig. 1*a*. The critical temperature $T_c$ at $B = 0$ T is defined as the inflection point of $\rho(T)$.

**Table 1.**

Temperatures of occurrence of the spin density wave ($T_{SDW}$) and/or superconducting ($T_c$) transition in the Ba(Fe$_{1-x}$Co$_x$)$_2$As$_2$ series. The measurements were carried out between 1.5 K and 300 K and the "-" mark means that no trace of a transition was observed in this temperature range.

| $x$ | 0 | 0.045 | 0.06 | 0.244 |
|---|---|---|---|---|
| $T_{SDW}$ | 135 K | 65 K | - | - |
| $T_c$ ($B = 0$ T) | - | 24.8 K | 20.5 K | - |

Fig. 2 presents the temperature dependences of the transverse thermal conductivity at $B$ = 12.5 T for the Ba(Fe$_{1-x}$Co$_x$)$_2$As$_2$ series. This quantity is determined as $\kappa_{xy} = \nabla_y T / \nabla_x T \, \kappa_{yy}$, where $\kappa_{yy}$ is assumed to be equal to $\kappa_{xx}$, since the Co6 and Co24 single crystals are tetragonal, whereas Co0 and Co4 are twinned. This means that due to domain structure of the samples [27] the transport coefficients measured along the *ab*-plane in the orthorhombic phase represents averaged values between the crystallographic *a* and *b* directions.

Studies of $\kappa_{xy}$ provide a unique opportunity to gain insight into the thermal transport properties of the electronic system which in high-$T_c$ superconductors are usually hidden by the dominating phonon contribution [22,28]. The transverse thermal conductivity describes solely properties of the electronic system (including the sign of charge carriers), because movement of chargeless quasiparticles is, in general, unaffected by the magnetic field. The phonon and magnon Hall effects may be exemptions from this rule, but their occurrence is limited to specific conditions met for example in paramagnetic [29] or ferromagnetic [30] insulators.

There are some interesting features of the $\kappa_{xy}(T)$ data presented in Fig. 2, especially the variation of $\kappa_{xy}$ due to the formation of the SDW order in the Co0 parent compound. While the Hall coefficient in this sample drops suddenly below $T_{SDW} = 135$ K, its thermal counterpart, the transverse thermal conductivity, behaves differently. Namely, $\kappa_{xy}$ shows an initial rise, reaching zero around $T \approx 105$ K, followed by a steep decrease and then a minimum at $T \approx 35$ K. Substantial differences between $\kappa_{xy}(T)$ and $R_H(T)$, or being more precise, between the transverse thermal and electrical ($\sigma_{xy}$) conductivities can be evidence for multiband electronic transport. For a single band, $\sigma_{xy} = \rho_{xy} / (\rho_{xx} \rho_{yy} + \rho_{xy}^2)$, where $\rho_{xy} = R_H B$, and in our case $\rho_{xx} = \rho_{yy}$ as well as $\rho_{xx} \gg \rho_{xy}$; therefore $\sigma_{xy} / B = R_H / \rho_{xx}^2$. In the two-band approximation with electron-like (e) and hole-like (h) charge carriers, the total $\kappa_{xy}$ is a sum of $\sigma_{xy}^e$ and $\sigma_{xy}^h$ weighted by the respective Hall Lorenz numbers ($L_{xy}$):

$$\kappa_{xy} = \kappa_{xy}^e + \kappa_{xy}^h = (L_{xy}^e \sigma_{xy}^e + L_{xy}^h \sigma_{xy}^h) T (e/k_B)^2. \tag{1}$$

$L_{xy}$ is defined by the Hall version of the Wiedemann-Franz law [31]:

$$L_{xy} \equiv \frac{\kappa_{xy}}{\sigma_{xy}T}\left(\frac{e}{k_B}\right)^2. \qquad (2)$$

A discrepancy between $\kappa_{xy}(T)$ and $\sigma_{xy}(T)$ was also observed in the iron-pnictide CaFe$_2$As$_2$, where it was attributed to the anomalous behavior of the transport coefficients of conductive bands with very small Fermi energies [32]. In such a case $L_{xy}^h$ of a parabolic band and $L_{xy}^e$ of the Dirac cone formed in the SDW state, depend differently on temperature and cause the total Hall Lorenz number to be negative in a certain temperature region [32]. Interestingly, there is no sign of such a behavior in the SDW phase of Co-doped Co4. This agrees well with studies of another iron-based superconductor Eu(Fe$_{1-x}$Co$_x$)$_2$As$_2$, where the authors report significant differences of the transport coefficients in the SDW state between the parent compound and lightly cobalt doped samples [33]. This suggests that the electron-like charge carriers in Co4 no longer act like Dirac fermions.

The electronic thermal to charge transport ratio, expressed in Eq. 3 by the dimensionless $L_{xy}$ reveals intriguing results, when compared between samples. Figure 3 presents temperature dependences of $L_{xy}$ in the Ba(Fe$_{1-x}$Co$_x$)$_2$As$_2$ series studied. A prominent feature of all $L_{xy}(T)$ plots is a plateau in the high temperature limit, where $L_{xy}$ reaches $L_0 = \pi^2/3$. $L_0$ is called the Sommerfeld value and it is acquired by the Lorenz number [34] as well as the Hall Lorenz number [31], if charge carriers in the Fermi liquid are scattered elastically.

$L_{xy}(T)$ of overdoped Co24 behaves differently than that of the parent compound Co0 – there is no evident anomaly in the Hall Lorenz number, and the value of $L_{xy}$ is close to $L_0$ in the entire temperature range. This characteristic alone is surprising, since the Lorenz number (along with its Hall counterpart $L_{xy}$ [30]) in normal metals is suppressed in the intermediate temperature range due to inelastic scattering of charge carriers on phonons [35]. The absence of such a suppression means that charge carriers in Ba(Fe$_{1-x}$Co$_x$)$_2$As$_2$ are scattered mostly elastically at all temperatures, perhaps due to weak electron-phonon coupling [36]. A small residual resistivity ratio (RRR), which for all cobalt doped samples is around 2, suggests scattering by Co atoms as a good candidate for the dominating elastic process.

Despite the fact that electron scattering in Ba(Fe$_{1-x}$Co$_x$)$_2$As$_2$ seems to be mostly elastic, $L_{xy}(T)$ in both Co4 and Co6 $L_{xy}$ rises with decreasing temperature, to reach a maximum at $T^{max} \approx 35$ K, $L_{xy}^{max} \approx 9.5$ and $T^{max} \approx 45$ K, $L_{xy}^{max} \approx 7$, respectively, then drops back to $L_0$ in the low temperature limit. An emergence of the maximum seems to be independent of the presence of the SDW state, which is absent in Co6, while in Co4 the onset of the maximum in $L_{xy}(T)$ occurs at a temperature much higher than $T_{SDW} = 65$ K. Namely, the value of the Hall Lorenz number starts to rise above the Sommerfeld value below $T^* \approx 140$ K for Co4 and $T^* \approx 125$ K for Co6. Figure 4 shows where the characteristic temperatures of the studied samples locate in the $x$ - $T$ phase diagram of Ba(Fe$_{1-x}$Co$_x$)$_2$As$_2$.

The maxima of the Hall Lorenz number in Co4 and Co6, where values of $L_{xy}$ exceed $L_0$, indicate a violation of WFL. This is clear evidence for unusual properties of the electronic system in Ba(Fe$_{1-x}$Co$_x$)$_2$As$_2$. A similar disturbance of the heat to charge transport ratio was already observed in chromium [37], where it was attributed to the anomalous distribution of the electronic states around the Fermi level [20, 21]. An analogous mechanism caused by the pseudogap was suggested as the origin of the enhanced high temperature part of the Lorenz number in YBa$_2$Cu$_3$O$_{7-y}$ [22] and the Hall Lorenz number in YBa$_2$Cu$_3$O$_{7-y}$[23] and EuBa$_2$Cu$_3$O$_{7-y}$ [24]. This phenomena can probably be understood in simple terms within Boltzmann transport theory. Formulas for transport coefficients are derived from the Boltzmann equation [38]:

the longitudinal electrical conductivity:

$$\sigma_{xx} = 2e^2 \sum_{k} \left(v_k^x\right)^2 \tau_k \left(-\frac{\partial f_k}{\partial \varepsilon_k}\right), \tag{3}$$

the longitudinal thermal conductivity:

$$\kappa_{xx} = \frac{2}{T} \sum_k \varepsilon_k^2 (v_k^x)^2 \tau_k \left( -\frac{\partial f_k}{\partial \varepsilon_k} \right), \quad (4)$$

the transverse electrical conductivity:

$$\sigma_{xy} = -2 \frac{e^3 B}{\hbar c} \sum_k \tau_k v_k^x \left[ v_k^y \partial_{k_x} + v_k^x \partial_{k_y} \right] \tau_k v_k^y \left( -\frac{\partial f_k}{\partial \varepsilon_k} \right), \quad (5)$$

and the transverse thermal conductivity:

$$\kappa_{xy} = -2 \frac{eB}{T\hbar c} \sum_k \varepsilon_k^2 \tau_k v_k^x \left[ v_k^y \partial_{k_x} + v_k^x \partial_{k_y} \right] \tau_k v_k^y \left( -\frac{\partial f_k}{\partial \varepsilon_k} \right), \quad (6)$$

where: $\varepsilon_k$ is the energy of the quasiparticles, $\tau_k$ is the relaxation time, $v_k$ is the group velocity of the quasiparticles, and $f_k$ is the Fermi distribution function. The only difference in the expressions under the summation sign for the electrical and thermal conductivity tensors is the $\varepsilon_k^2$ term present in the latter. A depletion in the electronic density of states near the Fermi level, or in other words the pseudogap, affects the electrical conductivity more severely, since states further away from the Fermi energy are more important for the thermal conductivity due to the presence of the $\varepsilon_k^2$ weighting factor [22]. This causes the Lorenz number and the Hall Lorenz number to rise above $L_0$. Moreover, because $\kappa_{xy}$ incorporates the sign of the dominant charge carriers one may conclude that the pseudogap in Ba(Fe$_{1-x}$Co$_x$)$_2$As$_2$ opens in the electron band. This is in agreement with results of angle-resolved photoemission spectroscopy studies of Ba$_{1-x}$K$_x$Fe$_2$As$_2$, which shows that PG opens in one particular sheet of the Fermi surface that seems to be the electron-like α band [15].

Solid lines in Fig. 3*a* (Co4) and 3*b* (Co6) show numerical calculations of $L_{xy}(T)$ using equations 2 – 6, where $\tau_k$ is assumed to be constant. The pseudogap is modeled as a quadratic depression of the electronic density of states at the Fermi level [39] that emerges below $T^*$. Its depth increases linearly to reach a constant value at temperatures just above the maximum in $L_{xy}(T)$. While the proposed model is very basic, it reproduces the experimental data quite well.

In addition, since the temperature of the maximum ($T^{max}$) in $L_{xy}(T)$ depends mostly on the width of the pseudogap, this can be estimated from $T^{max}$. Thus, the results suggest the presence of rather narrow pseudogaps, i.e. $\Delta_{PG} = 34$ meV for Co4 and $\Delta_{PG} = 30$ meV for Co6. These values are smaller than those deduced from the infrared spectroscopy, where $\Delta_{PG} = 87$ meV [40], while they are similar to the results of time resolved, optical pump probe spectroscopy studies: $\Delta_{PG} = 52 - 69$ meV [41] and comparable to the widths resulting from the analysis of NMR data: $\Delta_{PG} = 42 - 61$ meV [42]. On the other hand, the relation between the width of the pseudogap and the SC critical temperature, which is 16 $k_B T_c$ in Co4 and 17 $k_B T_c$ for Co6, fits well in the range $9 - 46$ $k_B T_c$ suggested for the "1111" iron-based superconductors [43], which in turn is also very similar to the $10 - 40$ $k_B T_c$ range observed in the cuprate superconductor [44].

**Summary**

In summary, we have comprehensively investigated heat and charge transport in the normal state for single crystals of the iron-based pnictide Ba(Fe$_{1-x}$Co$_x$)$_2$As$_2$. We focus on the transverse thermal and electrical conductivities that allow one to determine the Hall Lorenz number and to study the Wiedemann-Franz law. We find this fundamental relation fulfilled for all samples in the high and low temperature limits, which shows that quasiparticles in Ba(Fe$_{1-x}$Co$_x$)$_2$As$_2$ form the well defined Fermi liquid. Furthermore, the value of $L_{xy}$ seems not to be affected by inelastic scattering, even in the intermediate temperature region. This is probably because of relatively weak electron-phonon coupling combined with strong elastic scattering of the charge carriers by Co atoms. In the two superconducting samples we observe a violation of WFL, which can be ascribed to the emergence of a rather narrow pseudogap in the electron-like band. The temperatures below which $L_{xy}$ departs from its canonical Sommerfeld value correlate with the temperatures of the pseudogap formation known from other studies. Interestingly, the anomaly in $L_{xy}(T)$ seems to be independent of the presence of the spin density wave state.

**Acknowledgments**

The authors would like to thank J.R. Cooper for helpful comments.

This work was supported financially by the National Science Centre (Poland) under the research Grant No. 2011/03/B/ST3/00477.

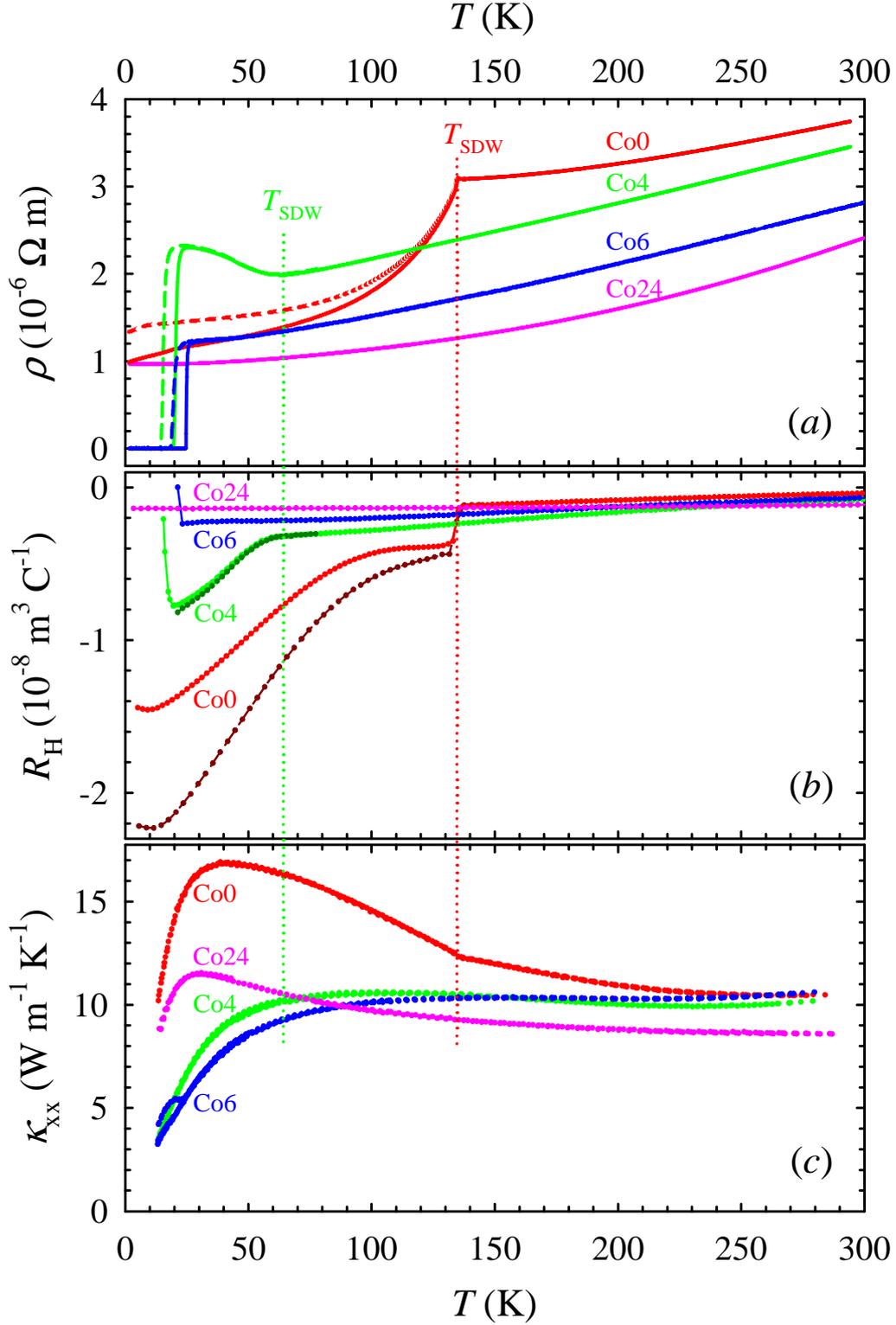

**Figure 1.**
(Color online) The temperature dependences of $\rho$, $R_H$, and $\kappa_{xy}$ for the Ba(Fe$_{1-x}$Co$_x$)$_2$As$_2$ series. In panel *a* solid lines represent data measured in $B = 0$ T, whereas dashed lines data are those measured in $B = 12.5$ T. In panel *b* solid lines are data for $B = 12.5$ T, and dark dashed lines for $B = 7$ T. The only noticeable difference between $B = 0$ T and $B = 12.5$ T data for the thermal conductivity in panel *c* is a small increase in zero field at $T_c$ of Co6. Dotted vertical lines mark $T_{SDW}$ in Co0 and Co4.

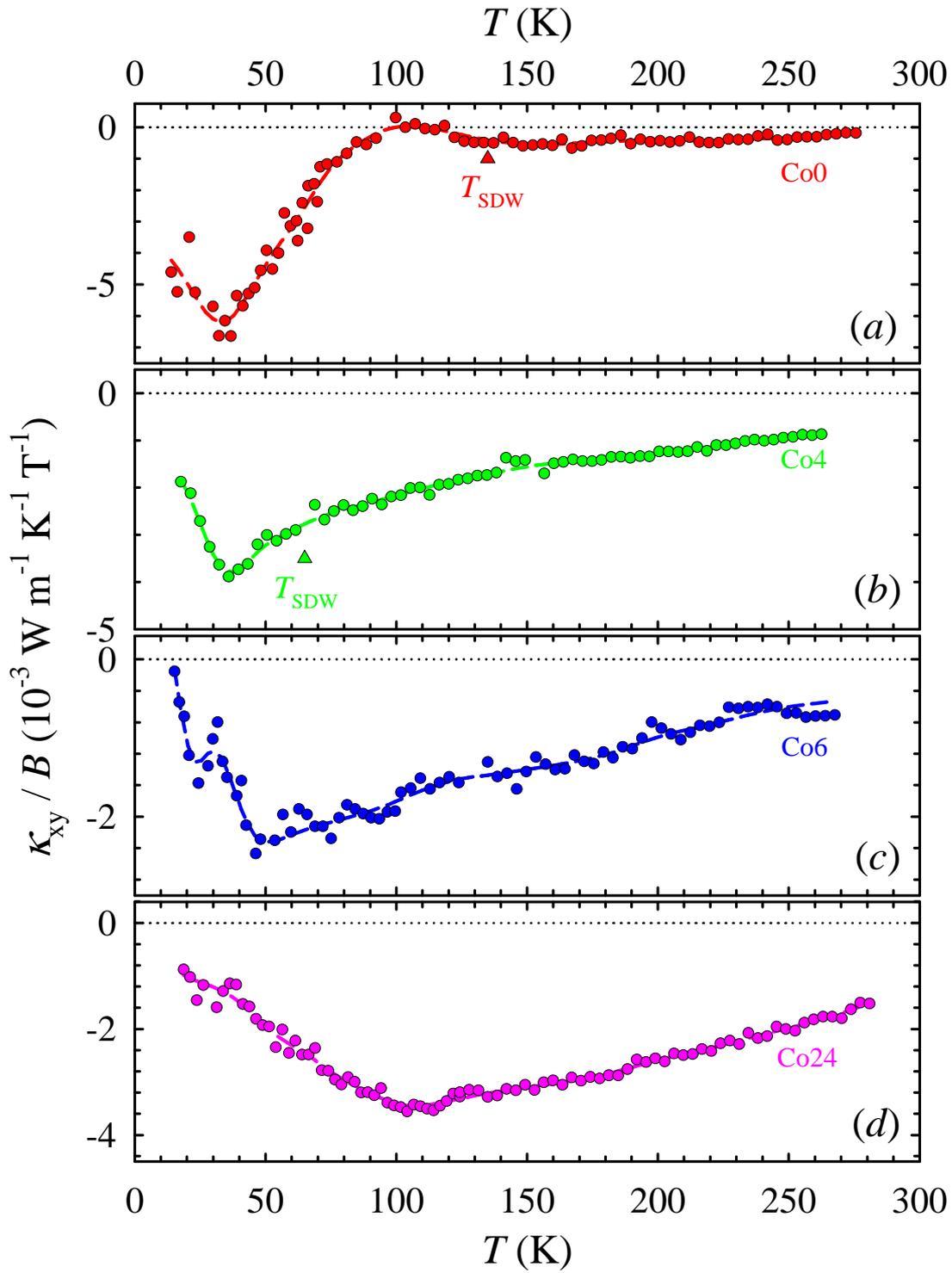

**Figure 2.**

(Color online) The temperature dependences of $\kappa_{xy}$ in $B = 12.5$ T for the Ba(Fe$_{1-x}$Co$_x$)$_2$As$_2$ series. Triangles in panels *a* and *b* indicate $T_{SDW}$. Dashed lines are guides for the eye.

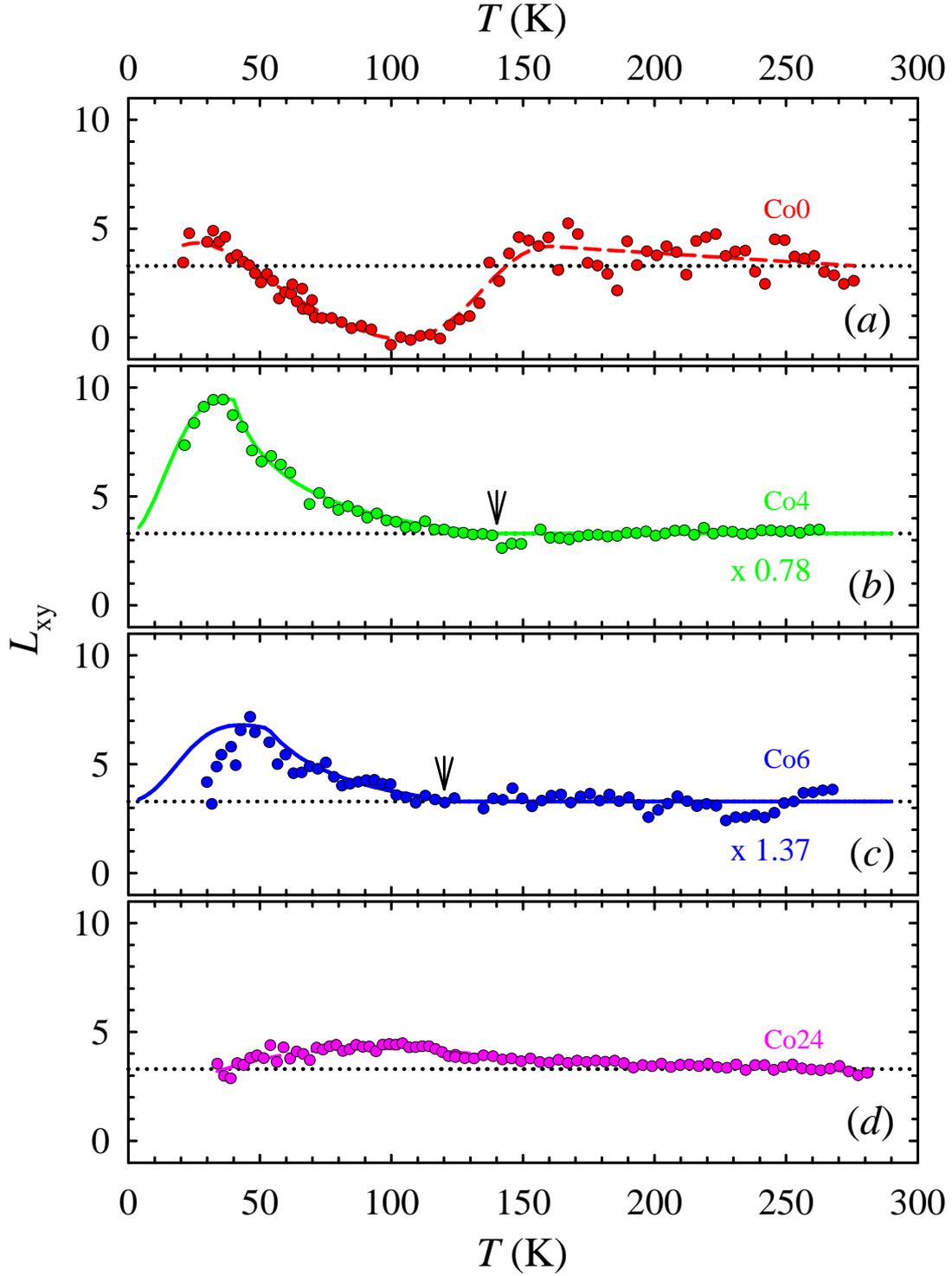

**Figure 3.**
(Color online) The temperature dependences of the Hall Lorenz number for the Ba(Fe$_{1-x}$Co$_x$)$_2$As$_2$ series. Arrows in panels *b* and *c* indicate onset of the maximum in $L_{xy}(T)$. Dashed lines in panels *a* and *d* are guides for the eye, whereas solid lines in panels *b* and *c* are curves calculated from the model described in the text. Dotted horizontal lines mark the Sommerfeld value $L_0 = \pi^2/3$. Numbers in panel (*b*) and (*c*) denote correcting constants close to 1 to cancel uncertainties related to measurement errors in geometric factors.

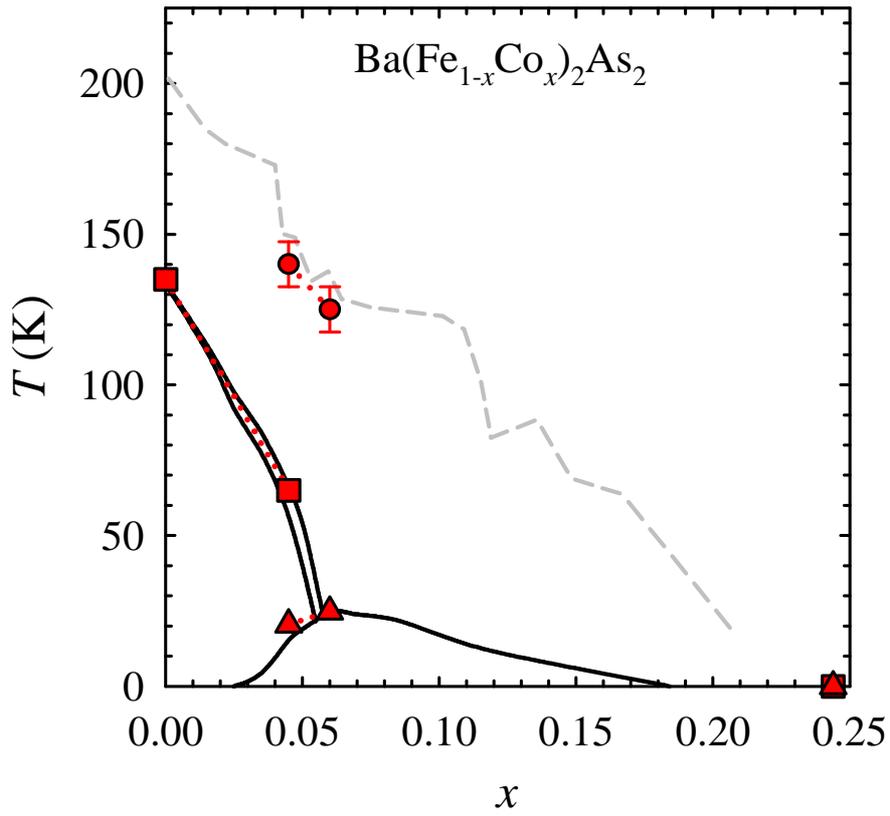

**Figure 4.**
(Color online) The *x-T* phase diagram of Ba(Fe$_{1-x}$Co$_x$)$_2$As$_2$. The solid lines denote the SDW, structural and SC transitions (reproduced from Ref.[45]), whereas the dashed line the proposed emergence of the pseudogap (from Ref. [17]). Points, squares and triangles show respective temperatures determined in the present work. Error bars denote uncertainty in determining the onset of the maximum in $L_{xy}(T)$.

**References**

[1] C. Renner, B. Revaz, J.Y. Genoud, K. Kadowaki and O. Fischer, Phys. Rev. Lett., 80, 149 (1998).

[2] O. Tjernberg, H. Nylen, G. Chiaia, S. Soderholm, U.O. Karlsson, M. Qvarford, I. Lindau, C. Puglia, N. Martensson and L. Leonyuk, Phys. Rev. Lett. 79, 499 (1997).

[3] H. Ding, T. Yokoya, J.C. Campuzano, T. Takahashi, M. Randeria, M.R. Norman, T. Mochiku, K. Kadowaki and J. Giapintzakis, Nature 382, 51 (1996).

[4] C.M. Varma, Phys. Rev. Lett. 83, 3538 (1999).

[5] J.L. Tallon, J.W. Loram, G.V.M. Williams, J.R. Cooper, I.R. Fisher, J.D. Johnson, M.P. Staines and C. Bernhard, Phys. Status Solidi b 215, 531 (1999).

[6] J.L. Tallon and J.W. Loram, Physica C 349, 53 (2001).

[7] D. Senechal and A.M.S. Tremblay, Phys. Rev. Lett. 92, 126401 (2004).

[8] S. Kawasaki, T. Mito, G.-q. Zheng, C. Thessieu, Y. Kawasaki, K. Ishida, Y. Kitaoka, T. Muramatsu, T.C. Kobayashi, D. Aoki, S. Araki, Y. Haga, R. Settai and Y. Onuki, Phys. Rev. B 65, 020504 (R) (2001).

[9] S. Sasaki, M. Kriener, K. Segawa, K. Yada, Y. Tanaka, M. Sato, and Y. Ando, Phys. Rev. Lett. 107, 217001 (2011).

[10] I. Pallecchi, M. Tropeano, C. Ferdeghini, G. Lamura, A. Martinelli, A. Palenzona, M. Putti, J. Supercond. Nov. Magn. 24, 1751 (2011).

[11] T. Sato, S. Souma, K. Nakayama, K. Terashima, K. Sugawara, T. Takahashi, Y. Kamihara, M. Hirano, and H. Hosono, Journal of the Physical Society of Japan 77, 063708 (2008).

[12] C. Hess, A. Kondrat, A. Narduzzo, J. E. Hamann-Borrero, R. Klingeler, J. Werner, G. Behr and B. Buchner, EPL 87, 17005 (2009).

[13] H.Y. Liu, X.W. Jia, W.T. Zhang, L. Zhao, J.Q. Meng, G.D. Liu, X.L. Dong, G. Wu, R.H. Liu, X.H. Chen, Z.A. Ren, W. Yi, G.C. Che, G.F. Chen, N.L. Wang, G.L. Wang, Y. Zhou, Y. Zhu, X.Y. Wang, Z.X. Zhao, Z.Y. Xu, C.T. Chen, X.J. Zhou, Chinese Phys. Lett. 25, 3761 (2008).

[14] T. Mertelj, V.V. Kabanov, C. Gadermaier, N.D. Zhigadlo, S. Katrych, J. Karpinski, and D. Mihailovic, Phys. Rev. Lett. 102, 117002 (2009).

[15] Y.M. Xu, P. Richard, K. Nakayama, T. Kawahara, Y. Sekiba, T. Qian, M. Neupane, S. Souma, T. Sato, T. Takahashi, H.Q. Luo, H.H. Wen, G.F. Chen, N.L. Wang, Z. Wang, Z. Fang, X. Dai and H. Ding, Nature Communications 2, 392 (2011).

[16] Y.S. Kwon, J.B. Hong, Y.R. Jang, H.J. Oh, Y.Y. Song, B.H. Min, T. Iizuka, S. Kimura, A.V. Balatsky, Y. Bang, New J. Phys. 14, 063009 (2012).

[17] M.A. Tanatar, N. Ni, A. Thaler, S.L. Budko, P.C. Canfield and R. Prozorov, Phys. Rev. B 82, 134528 (2010).

[18] S.J. Moon, A.A. Schafgans, S. Kasahara, T. Shibauchi, T. Terashima, Y. Matsuda, M.A. Tanatar, R. Prozorov, A. Thaler, P.C. Canfield, A.S. Sefat, D. Mandrus and D. N. Basov, Phys. Rev. Lett. 109, 027006 (2012).

[19] F. Massee, Y.K. Huang, J. Kaas, E. van Heumen, S. de Jong, R. Huisman, H. Luigjes, J.B. Goedkoop and M.S. Golden, EPL 92, 57012 (2010).

[20] J.F. Goff, Phys. Rev. B 1, 1351 (1970).

[21] J.F. Goff, Phys. Rev. B 2, 3606 (1970).

[22] H. Minami, V.W. Wittorff, E.A. Yelland, J.R. Cooper, C. Changkang and J.W. Hodby, Phys. Rev. B 68, 220503(R) (2003).

[23] M. Matusiak, K. Rogacki, B. W. Veal, EPL 88, 47005 (2009).

[24] M. Matusiak and T. Wolf, Phys. Rev. B 72, 054508 (2005).

[25] M. Matusiak, J. Hori, T. Suzuki, Solid State Communications 139 (2006) 376.

[26] F. Hardy, P. Adelmann, Th. Wolf, H.v. Löhneysen, C. Meingast, Phys. Rev. Lett. 102, 187004 (2009).

[27] R. Prozorov, M.A. Tanatar, N. Ni, A. Kreyssig, S. Nandi, S.L. Budko, A.I. Goldman, and P.C. Canfield, Phys. Rev. B 80, 174517 (2009).

[28] P.B. Allen, X. Du, L. Mihaly and L. Forro, Phys. Rev. B 49, 9073 (1994).

[29] C. Strohm, G.L.J.A. Rikken and P. Wyder, Phys. Rev. Lett. 95, 155901 (2005).

[30] Y. Onose, T. Ideue, H. Katsura, Y. Shiomi, N. Nagaosa, Y. Tokura, Science 329, 297 (2010).

[31] Y. Zhang, N.P. Ong, Z.A. Xu, K. Krishana, R. Gagnon and L. Taillefer, Phys. Rev. Lett. 84, 2219 (2000).

[32] M. Matusiak, Z. Bukowski, and J. Mucha, JPS Conf. Proc. 3, 015002 (2014).

[33] M. Matusiak, Z. Bukowski, and J. Karpinski, Phys. Rev. B 83, 224505 (2011).

[34] N.W. Ashcroft and N.D. Mermin, Solid State Physics (Harcourt College Publishers, Fort Worth, 1976).

[35] J.M. Ziman, Electrons and Phonons: The Theory of Transport Phenomena in Solids (Clarendon Press, Oxford, 1960).

[36] B. Mansart, D. Boschetto, A. Savoia, F. Rullier-Albenque, F. Bouquet, E. Papalazarou, A. Forget, D. Colson, A. Rousse, and M. Marsi, Phys. Rev. B 82, 024513 (2010).

[37] G.T. Meaden, K.V. Rao, and H.Y. Loo, Phys. Rev. Lett. 23, 475 (1969).

[38] A. Perali, M. Sindel, and G. Kotliar, Eur. Phys. J. B 24, 487 (2001).

[39] J. Hwang, J. Yang, T. Timusk, S. G. Sharapov, J. P. Carbotte, D.A. Bonn, R.X. Liang, and W.N. Hardy, Phys. Rev. B 73, 014508 (2006).

[40] S.J. Moon, Y.S. Lee, A.A. Schafgans, A.V. Chubukov, S. Kasahara, T. Shibauchi, T. Terashima, Y. Matsuda, M.A. Tanatar, R. Prozorov, A. Thaler, P.C. Canfield, S.L. Budko, A.S. Sefat, D. Mandrus, K. Segawa, Y. Ando, D.N. Basov, Phys. Rev. B 90, 014503 (2014).

[41] L. Stojchevska, T. Mertelj, J. H. Chu, I.R. Fisher, and D. Mihailovic, Phys. Rev. B 86, 024519 (2012).

[42] F.L. Ning, K. Ahilan, T. Imai, A.S. Sefat, R.Y. Jin, M.A. McGuire, B.C. Sales, and D. Mandrus, J. Phys. Soc. Jpn. 78, 013711 (2009).

[43] H.W. Ou, Y. Zhang, J.F. Zhao, J. Wei, D.W. Shen, B. Zhou, L.X. Yang, F. Chen, M. Xu, C. He, R.H. Liu, M. Arita, K. Shimada, H. Namatame, M. Taniguchi, Y. Chen, X.H. Chen, D.L. Feng, Solid State Commun. 148, 504 (2008).

[44] A. Damascelli, Z. Hussain, Z.-X. Shen, Rev. Mod. Phys. 75 (2003) 473.

[45] J.-H. Chu, J.G. Analytis, C. Kucharczyk, and I.R. Fisher, Phys. Rev. B 79, 014506 (2009).